\documentclass[12pt]{article}

\def\s{\sigma}
\def\be{\begin{equation}}
\def\ee{\end{equation}}
\def\LL{${\cal L}$ }
\def\N{${\cal N}$ }
\def\2{\sqrt 2}
\def\3{\sqrt 3}
\def\a{\alpha}
\def\b{\beta}
\def\d{\delta}
\def\e{\epsilon}
\def\t{\theta}
\def\l{\lambda}
\def\bea{\begin{eqnarray}}
\def\eea{\end{eqnarray}}

\begin{document}

 \title{Theory of impedance networks: \\The two-point impedance and $LC$ resonances}
 \vskip 1cm
\author{W. J. Tzeng\\ Department of Physics\\
 Tamkang University,  Taipei, Taiwan\\
and \\
F. Y. Wu \\ Department of Physics \\
Northeastern University, Boston, Massachusetts 02115, U.S.A.}

\date{}
\maketitle

\abstract{We present a formulation of the determination of the impedance between
any two nodes in an impedance network. An impedance network is described
by its Laplacian matrix ${\bf L}$ which has  generally complex matrix elements.
We show that by solving
the equation  ${\bf L}\, u_\a = \lambda_\a \,u_\a^*$ with orthonormal
  vectors $u_a$, the effective impedance between nodes $p$ and $q$ of the network is
$Z_{pq} = \sum_{\a}
(u_{\a\,p} - u_{\a\,q})^2/{\lambda_\a}$
where the summation is over all $\lambda_\a$ not identically equal to zero
and $u_{\a\, p}$ is the $p$-th component of $u_\a$.
For networks consisting of inductances $L$ and capacitances $C$,
the formulation leads to the occurrence of resonances at
 frequencies associated with
the vanishing of $\lambda_\a$.
This curious result suggests the possibility of practical applications
to resonant circuits.
   Our formulation is illustrated by explicit
examples.}

 \vskip 1cm



 \newpage

\section{Introduction}
A classic problem in electric circuit theory that has attracted attention
from Kirchhoff's time \cite{kirch}  to the present
is the consideration
of network resistances and impedances.
While the evaluation of resistances and impedances can in principle be carried out
for any given network using  traditional, but often tedious,
analysis such as the Kirchhoff's laws, there has been no
conceptually simple solution.  Indeed, the problem of computing
the effective resistance between
two arbitrary nodes in a resistor network has been studied by numerous authors (for a list of
relevant references on resistor networks up to 2000 see, e.g., \cite{cserti}).
Particularly, an elementary exposition of the material can be found
in Doyle and Snell \cite{ds}.

\medskip
 However, past efforts  prior to 2004 have been focused
mainly on regular lattices and  the use of the Green's function technique,
for which the analysis is most conveniently carried out when the network size is
 infinite \cite{cserti,cserti1}.   Little attention has been
paid to {\it finite} networks, even though the latter are those occurring in applications.
Furthermore, there has been very few studies on
impedance networks.
To be sure, studies have been carried out on electric and optical properties of
random impedance networks in binary composite media (for a review see \cite{clerc})
and in dielectric resonances occurring
in clusters embedded in a regular lattice \cite{clerc1}. But these are
mostly approximate treatments on random media.  More recently Asad et al.
\cite{asad}  evaluated the two-point capacitance in an infinite network
of identical capacitances.  When all impedances in a network
are identical, however, the Green's function technique used
   and the results are essentially the same as
  those of identical  resistors.

\medskip
In 2004 one of us proposed\cite{wu} a new formulation of resistor networks which leads to
an expression of  the effective resistance between any two nodes in a
network in terms of
the eigenvalues and eigenvectors of the Laplacian matrix.
Using this formulation one  computes the effective resistance between
two arbitrary nodes in any network which can be either finite or infinite \cite{wu}.
 This is a fundamentally new formulation.
But the analysis presented in \cite{wu} makes
use of the fact that,  for resistors the Laplacian matrix has real matrix elements.
Consequently the method does not extend
to impedances whose  Laplacian matrix elements are   generally complex
(see e.g., \cite{2}).
  In this paper we resolve this difficulty and extend the formulation of \cite{wu} to impedance
networks.

\medskip
Consider an impedance network \LL consisting of \N nodes numbered $\a=1,2,...,{\cal N}$.
Let the impedance connecting nodes $\a$ and $\b$ be
\be
z_{\a\b}=z_{\b\a} = r_{\a\b} + i\, x_{\a\b} \label{imp}
\ee
  where $r_{\a\b}=r_{\b\a}\geq 0$ is the resistive part
and $x_{\a\b}=x_{\b\a}$ the reactive part, which is positive for inductances
and negative for capacitances.  Here, $i=\sqrt {-1}$
often denoted by $j=\sqrt{-1}$ in alternating current (AC) circuit theory \cite{2}.
In this paper we shall use $i$ and $j$ interchangeably.
 The admittance $y$ connecting two nodes is the reciprocal of the
 impedance.  For example,
 $y_{\a\b}=
y_{\b\a}=1/z_{\a\b}$.

\medskip
Denote the electric potential at node $\a$  by $V_\a$ and the {\it
net} current flowing into the network (from the outside world) at
node $\a$ by $I_\a$\, . Both  $V_\a$ and $I_\a$ are generally
complex in the {\it phasor} notation used in AC circuit theory \cite{2}.
Since there is neither source nor
 sink of currents,   one has
the conservation rule
\be
 \sum_{\a=1}^{\cal N} I_\a =0\, .
\label{conserve}
\ee
 The Kirchhoff equation for the network reads
 \be
 {\bf
L}\, {\vec V} = {\vec I} \label{Kirchhoff} \ee
where
 \bea
 {\bf L}
&=& \pmatrix {y_1&-y_{12}& \dots &\ -y_{1{\cal N}}\cr
                    -y_{21} &\ y_2 & \dots & -y_{2{\cal N}} \cr
                     \vdots & \vdots & \ddots & \vdots & \cr
   -y_{{\cal N}1} & -y_{{\cal N}2} & \dots &\ y_{{\cal N}} \cr},
\label{Kmatrix} \eea
 with
\bea y_\a \equiv \sum^{\cal N}_{\b=1\,(\b\not=\a)} \
y_{\a\b},
\label{Xdef} \eea is the  Laplacian matrix associated with
the network ${\cal L}$. In (\ref{Kirchhoff}),  ${\vec V}$ and ${\vec
I}$ are ${\cal N}$-vectors whose
 components are respectively $V_\a$ and $I_\a$.

\medskip
Here, we need to solve (\ref{Kirchhoff}) for ${\vec V}$ for a
given current configuration ${\vec I}$.
The effective impedance between nodes $p$ and $q$, the quantity we wish to
compute, is by definition the ratio
 \be
 Z_{pq} =
{{V_p-V_q}\over I}\label{equalr}
 \ee
where $V_p$ and $V_q$ are solved from  (\ref{Kirchhoff})
 with
 \bea
I_\a =I(\d_{\a\, p} -\d_{\a\, q}) . \label{I}
\eea

\medskip
The crux of matter is to solve the Kirchhoff equation
(\ref{Kirchhoff}) for ${\vec I}$ given by (\ref{I}). The difficulty lies
in the fact that, since the matrix {\bf L} is singular,
the equation (\ref{Kirchhoff}) cannot be
formally inverted.

\medskip
To circumvent this difficulty we proceed as in \cite{wu} to consider
instead the equation \be {\bf L}(\e)\, {\vec V}(\e) = {\vec I}
\label{modKirchhoff} \ee where \be {\bf L}(\epsilon) ={\bf L}
+\epsilon \,{\bf I} \label{modifiedL} \ee
 and {\bf I} is the identity
matrix. The matrix ${\bf L}(\e)$ now has an inverse and we can
proceed by applying the arsenal of linear algebra.
 We  take  the
 $\e \to 0$ limit at the end and do not expect any problem
since we know there is a physical solution.

\medskip
The crucial step is the computation of the inverse matrix ${\bf
L}^{-1}(\e)$.
For this purpose it is useful to first recall the approach for resistor networks.

\medskip
In the case of resistor networks the matrix ${\bf L}(\epsilon)$ is
real symmetric and hence it has orthonormal
eigenvectors $\psi_\a(\epsilon)$ with eigenvalues
$\lambda_\a(\epsilon)=\lambda_\a +\epsilon$ determined from the
eigenvalue equation
\be {\bf L}(\e)\, \psi_\a(\e) = \lambda_\a(\e)
\,\psi_\a(\e),\quad i=1,2,...,{\cal N}.\label{Reigen}
\ee
Now a real hermitian matrix
${\bf L}(\e)$ is diagonalized by the unitary transformation ${\bf U}^\dagger(\e)\,
 {\bf L}(\e)\,{\bf U}(\e)=\Lambda(\e)$,
where  ${\bf U}(\epsilon)$ is a unitary matrix
 whose columns are the orthonormal eigenvectors
$\psi_\a(\epsilon)$ and $\Lambda(\e)$ is a diagonal matrix with
diagonal elements $\lambda_\a(\e)=\lambda_\a + \e $. The inverse of
this relation leads to ${\bf L}^{-1}(\e)={\bf
U}(\e)\,\Lambda^{-1}(\e) \, {\bf U}^{\dagger}(\e)$.
{\footnote {The equivalent of the method we use in obtaining (\ref{II}) is known
in mathematics literature
as the  pseudo-inverse method (see, e.g., \cite{inverse,inverse1})}}
In this way we find
the effective resistance between nodes $p$ and $q$ to be \cite{wu}
\be
 R_{pq} =
\sum_{\a=2}^{\cal N} \frac 1 {\lambda_\a} \big| \psi _{\a\,p} -
\psi _{\a\,q} \big|^2 \label{II}
\ee
where the summation is over
all nonzero eigenvalues, and $\psi_{\a\,p}$ is the
$p$th component of $\psi_\a(0)$. Here the $\a=1$ term in the summation
with $\lambda_1(\e)=\epsilon$ and $\psi_{1\,p}(\e)=1/\sqrt{\cal N}$
drops out (before taking the $\e\to 0$ limit) due to the
conservation rule (\ref{conserve}). It can be shown that there is no
other zero eigenvalue if the network is singly-connected.
The relation (\ref{II}) is the main result of \cite{wu}.

 \section{Impedance networks}
For impedance networks the Laplacian matrix {\bf L} is symmetric
and generally complex  and thus
 \bea
  {\bf L}^\dagger =
{\bf L}^* \not= {\bf L}\nonumber
\eea
where $*$ denotes the complex conjugation
and $\dagger$ denotes the hermitian conjugation.
Therefore ${\bf L}$ is not hermitian and cannot be
diagonalized as described in the preceding section.

\medskip
However, the matrix
  ${\bf L}^\dagger\,{\bf L}$ is always hermitian and has nonnegative
eigenvalues.  Write the eigenvalue equation as
\be
{\bf L}^\dagger\,{\bf L} \, \psi_\a = \s_\a\, \psi_\a\ , \quad \s_\a \geq 0, \quad \a=1,2,...,{\cal N}.\label{eigen}
\ee
One verifies that one eigenvalue is $\s_1 = 0$
with $\psi_1=\{1,1,...,1\}^T/\sqrt{\cal N}$, where the superscript $T$ denotes
the transpose.  For complex {\bf L}
 there can exist other zero eigenvalues (see below).

\medskip
To facilitate  considerations, we again introduce ${\bf L}(\e)$
as in (\ref{modifiedL}) and rewrite ({\ref{eigen}) as
\be
{\bf L}^\dagger(\e)\,{\bf L} (\e)\, \psi_\a(\e) = \s_\a(\e) \psi_\a (\e)\
, \quad \s_\a(\e) \geq 0, \quad \a=1,2,...,{\cal N},
\label{modifiedeigen}
\ee
where $\e$ is small.
 Now one eigenvalue is $\s_1(\e) = \e^2$
with $\psi_1(\e) =\{1,1,...,1\}^T/\sqrt{\cal N}$.
For other eigenvectors we make use of the theorem
established in the next section (see also \cite{horn}) that
 there exist ${\cal N}$ orthonormal vectors $u_\a(\e)$ satisfying the equation
  \be
{\bf L}(\e)\, u_\a(\e) = \lambda_\a(\e) \,u_\a^*(\e),\quad a=1,2,...,{\cal N}\label{modZeigen}
\ee
where
\be
\lambda_\a(\e) = \sqrt{\s_\a(\e)}\,e^{i\theta_\a(\e)}, \quad \theta_\a(\e) = {\rm real}.
\ee
Particularly,  we can take
\be
\lambda_1(\e) = \sqrt{\s_1(\e)} = \e, \quad \theta_1(\e) = 0.
\ee
Equation (\ref{modZeigen}) plays the role of the eigenvalue
equation (\ref{Reigen}) for resistors.

\medskip
   We next  construct a unitary matrix ${\bf U}(\e)$
whose columns are $u_\a(\e)$. Using (\ref{modZeigen}) and the fact that
${\bf L}(\e)$ is symmetric, one verifies that ${\bf L}(\e)$ is diagonalized
by the transformation
\bea
{\bf U}^T (\e)\, {\bf L}(\e)\, {\bf U}(\e) = \Delta (\e) \nonumber
\eea
where
$\Delta(\e)$ is a diagonal matrix with diagonal elements ${\lambda_\a(\e)}$.
The inverse of this relation leads to
 \be
 {\bf L}^{-1}(\e)  = {\bf U}(\e) \Delta^{-1}(\e) {\bf U}^T (\e). \label{modinverse}
\ee
where
$\Delta^{-1}(\e)$ is a diagonal matrix with diagonal elements ${1/\lambda_\a(\e)}$.
 We can now use (\ref{modinverse}) to solve
 (\ref{modKirchhoff}) to obtain,after using (\ref{equalr}),
\be
Z_{pq} = \lim _{\e \to 0}
 \sum_{\a=1}^{\cal N} \frac 1 {\lambda_\a(\e)}
\Big(u_{\a\,p}(\e) - u_{\a\,q}(\e)\Big)^2, \label{II1}
\ee
where $u_{\a\,p}$ is the $p$th component of the orthonormal vector $u_\a(\e)$.

\medskip
Now the term $\a=1$ in the summation drops out before taking the limit
just like in the case of resistors\cite{wu} since
 $\lambda_1(\e)=\e$ and  $u_{1\, p}(\e)=  u_{1\, q}(\e)=$ constant.
 If there exist other eigenvalues $\lambda_\a (\e)=\e$
with $u_{\a\, p}(\e) \not= $ constant, a situation
which can happen when there are pure reactances $L$ and $C$, the corresponding terms in (\ref{II1}) diverge
in the $\e\to 0$ limit at specific frequencies $\omega$ in an AC circuit.
 Then one obtains the  effective impedance
\bea
Z_{pq} &=& \sum_{\a=2}^{\cal N} \frac  1 {\lambda_\a }
\Big(u_{\a\,p} - u_{\a\,q}\Big)^2, \quad {\rm if\>\>}\lambda_\a\not= 0,\, \a \geq 2 \nonumber \\
  &=& \infty,  \hskip 1.4cm {\rm if\>\>there\>\>exists\>\>}\lambda_\a=0, \,\a\geq 2.\label{Impedance}
\eea
Here $u_{\a\,p}=u_{\a\,p}(0)$.
 The physical interpretation of $Z = \infty$
is the occurrence of  a
 {\it resonance}  in an AC circuit at frequencies where $\l_\a=0$,
 meaning it requires essentially a zero input current $I$ to maintain potential differences
at these frequencies.

\medskip
The expression (\ref{Impedance})  is our main result for impedance networks.

\medskip
In the case of pure resistors,  the Laplacian ${\bf L}(\e)$  and the
eigenvalues $\l_\a(\e)$ in  (\ref{Reigen}) are  real, so without the loss of
generality we can take  $\psi_\a(\e)$ to be real (see Example 3 in Sec. 5 below),
and use $u_\a(\e)= \psi_\a (\e)$ in (\ref{modZeigen}) with $\theta_\a(\e) =0$.
 Then $u_{\a \, p}(\e)$ in (\ref{II1}) is real and   (\ref{Impedance})
coincides with (\ref{II}) for resistors.  There is no
$\l_\a = 0$ other than $\l_1 = 0$, and there is no resonance.

\section{Complex symmetric matrix}
For completeness in this section we give a proof of
 the theorem which asserts (\ref{modZeigen})
and determines $u_\a$ for a complex symmetric matrix. Our proof parallels that
in \cite{horn}.

\medskip
\noindent
{\it Theorem:

\medskip
 Let {\bf L} be an $n\times n$ symmetric matrix with generally complex elements.
Write the eigenvalue equation of    ${\bf L}^\dagger\,{\bf L}$  as
 \be
{\bf L}^\dagger\,{\bf L} \, \psi_\a =
\s_\a\, \psi_\a \ , \quad \s_\a \geq 0, \quad \a=1,2,...,n. \label{Meigen}
\ee
  Then, there exist
$n$ orthonormal vectors \, $u_\a$ \,
  satisfying the relation
\be
{\bf L}\, u_\a =\lambda_\a \, u_\a^* ,\quad \a=1,2,...,n \label{u}
\ee
where $*$ denotes complex conjugation and $\lambda_\a= \sqrt{\s_\a}\, e^{i\theta_\a}$, $\theta_\a =$ real.

\medskip
 For nondegenerate $\s_\a$ we can take $u_\a=\psi_\a$;
for degenerate $\s_\a$, the $u$'s are linear combinations of the
degenerate $\psi_\a$.   In either case the phase factor $\theta_\a$
of $\lambda_\a$ is determined by applying (\ref{u}),}

\medskip
\noindent
Remarks:

\medskip
1. The  $\lambda_\a$'s are  the eigenvalues of ${\bf L}$ if $u_\a$'s
are real.

\medskip
2. If  \{$u_\a$, $\lambda_\a\}$ is a  solution of (\ref{u}), then
 $\{u_\a\,
e^{i\tau}, \lambda_\a\,e^{2i\tau}\}, \, \tau={\rm real}$, is also a solution
of (\ref{u}).

\medskip
3. While the procedure of constructing $u_\a$ in the
degenerate case appears to be involved, as
demonstrated in examples given in section 5  the orthonormal $u$'s can often be
determined quite directly in practice.

\medskip
4. If ${\bf L}$ is real, then as aforementioned
  it has real
eigenvalues and eigenvectors and
  we can take these real eigenvectors to be the $u_\a$ in (\ref{u})  with
$\l_\a$  real non-negative.

 \bigskip
 \noindent
Proof.

\medskip
Since   ${\bf L^\dagger\, L}$ is Hermitian its nondegenerate eigenvectors $\psi_\a$
can be chosen  to be orthonormal.
 For the eigenvector $\psi_\a$  with nondegenerate eigenvalue $\s_\a$,
construct a vector
\be
\phi_\a = \big( {\bf L} \psi_\a \big)^* +c_\a \psi _\a\,  \label{phia}
\ee
where $c_\a$ is any complex number.
It is readily verified that we have
\be
{\bf L}^\dagger\,{\bf L} \, \phi_\a = \s_\a\, \phi_\a\ , \label{Meigen1}
\ee
so  $\phi_\a$ is also an eigenvector of  ${\bf L}^\dagger\,{\bf L} $
 with the same eigenvalue $\s_\a$.
It follows that if $\s_\a$ is nondegenerate then $\phi_\a$ and
$\psi_\a$ must be proportional, namely, \be {\bf L}\, \psi_\a =
\lambda_\a \, \psi_\a^* \label{MM} \ee for some  $\lambda_\a$. The
substitution of (\ref{MM}) into (\ref{Meigen1})
with $\phi_\a$ given by (\ref{phia})
now yields
$|\lambda_\a|^2=\s_\a$ or $\lambda_\a = \sqrt{\s_\a}\, e^{i\theta_\a}$.
 Thus, for nondegenerate $\s_\a$ we simply choose $u_\a = \psi_\a$ and
use (\ref{u}) and (\ref{Meigen}) to determine the phase factor $\theta_\a$.
This  establishes the theorem for non-degenerate $\lambda_\a$.

\medskip
For degenerate  eigenvalues of ${\bf L^\dagger\, L}$, say, $\s_1=\s_2=\s $ with linearly independent
eigenvectors $\psi_1$ and $\psi_2$, we
construct
 \bea
&& v_1 =({\bf L}\, \psi_1)^* + \sqrt {\s} \, e^{i\theta_1}\,\psi_1\,   \nonumber \\
&& v_2 =({\bf L}\, \psi_2)^* + \sqrt{\s} \, e^{i\theta_2}\, \psi_2\,
\eea
where the choice of the real phase factors $\theta_1, \theta_2$ is at our disposal.
We choose $\theta_1, \theta_2$ to make $v_1$ and $v_2$ linearly independent
to satisfy
\be
e^{i(\theta_1 - \theta_2)} = (v_2,v_1)^*/(v_2,v_1) \label{v2v1}
\ee
where
$(y,z) = (y^T)^* z$ is the inner product of vectors $y$ and $z$.

\medskip
Now one has
\bea
{\bf L} \, v_1 &=& \sqrt {\s} \, e^{i\theta_1}\, v_1^*, \quad {\bf L} \, v_2 = \sqrt {\s} \,e^{i\theta_2}\,v_2^*\,\nonumber \\
{\bf L} ^\dagger\, {\bf L} \,v_1 &=& {\s} \, v_1, \hskip1.45cm {\bf L} ^\dagger\, {\bf L}\, v_2 =  {\s}\, v_2\,. \label{degenerate}
 \eea
Write
\be
u_1 = v_1/|v_1|\,,
\ee
where $|v| = \sqrt{ (v,v)}$ is the norm of $v$, and construct
$y = v_2 - (v_2, u_1) u_1$ which
 is orthogonal to $u_1$.
Next write
\be
u_2 = y/|y|.
\ee
Then, it can be verified by using (\ref{v2v1}) that $u_1$ and $u_2$ are orthonormal and satisfy
\bea
{\bf L} \, u_1 &=& \sqrt {\s} \, e^{i\theta_1}\, u_1^* \nonumber \\
{\bf L} \, u_2 &=& \sqrt {\s} \, e^{i\theta_2}\, u_2^*\, . \eea In
addition, both $u_1$ and $u_2$ are eigenvectors of ${\bf L}^\dagger\,{\bf L}$
with the same eigenvalue $\s$, hence are orthogonal to $\psi_\a, \,
\a\geq 3$. This establishes the theorem.

 \medskip
In the case of multi-degeneracy, a similar analysis can be carried out by
 starting from a set of  $v_\a$ to construct $u_\a$'s by using, say,  the
Gram-Schmidt orthonormalization procedure. For details we refer to \cite{Horn1}.

\section{Resonances}
If there exist eigenvalues $\lambda_\a = 0, \ \a \geq 2$, a situation which can occur
at specific frequencies $\omega$ in an AC circuit, then the effective impedance (\ref{Impedance})
between {\it any} two nodes
diverge and the network is  in  resonance.

\medskip
In an AC circuit resonances occur when
the impedances are pure reactances (capacitances or
inductances).  The simplest  example of a resonance
 is a circuit containing two nodes connecting
an inductance $L$ and capacitance $C$ in parallel.  It is well-known that
this $LC$ circuit is resonant with an external AC source at the frequency
$\omega = 1/\sqrt{LC}$. This is most simply seen by noting
that the two nodes are connected by an admittance $y_{12}=j\omega C + 1/j \omega L = j(\omega C -1/\omega L)$,
and hence $Z_{12}=1/y_{12}$ diverges at $\omega = 1/\sqrt{LC}$.

\medskip
Alternately, using our formulation, the Laplacian matrix is
\be
{\bf L} = y_{12} \pmatrix{1 & -1 \cr -1& 1}
\ee
so that ${\bf L}^*{\bf L}$ has eigenvalues $\s_1= 0,\, \s_2= 4|y_{12}|^2$
and we have $\lambda_1=0$ as expected.
In addition, we also have $\lambda_2 =0$
 when $y_{12}=0$ at the frequency $\omega = 1/\sqrt{LC}$.
 This is the occurrence of a resonance.

\medskip
An extension of this consideration to $N$
reactances in a ring is discussed in Example 2 in the next section.

\section{Examples}
\bigskip
Examples of applications of the formulation (\ref{Impedance}) is given in this
section.
\medskip

\noindent
{\it Example 1.  A numerical example.}

 \medskip
It is instructive to work out a numerical example as an illustration.

\medskip
Consider three impedances $z_{12}= i\sqrt 3,
z_{23}= -i\sqrt 3, z_{31} = 1$ connected in a ring as shown in Fig. 1
where $i=j=\sqrt {-1}$.
 We have the Laplacian
\be
{\bf L} = \pmatrix{1- i /{\sqrt 3} & i /{\sqrt 3} & -1 \cr
              i /{\sqrt 3} & 0 &   -i/ {\sqrt 3} \cr
              -1 & -  i/ {\sqrt 3} & 1+   i /{\sqrt 3}}.
\ee
Substituting {\bf L} into (\ref{eigen}) we find the following
nondegenerate eigenvalues and orthonormal eigenvectors of ${\bf L}^\dagger\, {\bf L}$,
\bea
\s_1 &=& 0, \hskip 1.6cm \psi_1 = \pmatrix{1\cr 1\cr 1}, \nonumber \\
\s_2 &=& 3-2\sqrt 2, \quad \psi_2 = \frac 1 {\sqrt{24-6\sqrt 2}}
    \pmatrix { 2-\sqrt 2 +i\sqrt 3 \cr -\sqrt 2 -1 - i\3 \cr 2\2 -1}, \nonumber \\
\s_3 &=& 3+2\2, \quad \psi_3 = \frac 1 {\sqrt{24+6\sqrt 2}}
    \pmatrix { 2+\sqrt 2 +i\sqrt 3 \cr \sqrt 2 -1 - i\3 \cr -2\2 -1}.
\eea
Since the eigenvalues are nondegenerate according to the theorem we take $u_i = \psi_i, i=1,2,3$.
Using these expressions we obtain from (\ref{u})
\bea
\sqrt {\s_2} &=& \2 -1, \quad e^{i\theta_2} = \frac 1 7
\Big[3\2-2 +i\3(2\2+1)\Big] \nonumber \\
\sqrt {\s_3} &=& \2 +1, \quad e^{i\theta_3} = \frac 1 7
\Big[3\2+2 +i\3(2\2-1)\Big] .
\eea
Now (\ref{Impedance}) reads
\be
Z_{pq} =  \frac { e^{-i\theta_2}} {\sqrt {\s_2}}
\Big(u_{2p} - u_{2q}\Big)^2 +  \frac { e^{-i\theta_3}} {\sqrt {\s_3}}
\Big(u_{3p} - u_{3q}\Big)^2  ,
\ee
using which one obtains the impedances
\be
Z_{12}=3+i\3, \quad Z_{23}=3-i\3, \quad Z_{31} =0.
\ee
These values agree with results of direct calculation using the Ohm's law.

\bigskip
\noindent
{\it Example 2. Resonance in a one-dimensional ring of $N$ reactances.}

\medskip
Consider $N$ reactances $jx_1, jx_2, ..., jx_N$ connected in a ring as shown in Fig. 2,
where $x=\omega L$ for inductance $L$ and $x=-1/\omega C$ for capacitance $C$ at AC frequency $\omega$.
The Laplacian assumes the form
\bea
{\bf L} =\frac 1 j  \left(\begin{array}{ccccccc}
y_1+y_N& -y_1&0&\cdots&0&0&-y_N\\
-y_1&y_1+y_2&-y_2 &\cdots&0&0&0\\
\vdots&\vdots&\vdots&\ddots&\vdots&\vdots&\vdots\\
0&0&0&\cdots&-y_{N-1}&y_{N-1}+y_N&-y_{N}\\
-y_N&0&0&\cdots&0&-y_1&y_N+y_1
\end{array}
\right), \label{Lper}
\eea
where $y_i=1/x_i$.
The Laplacian {\bf L} has one zero eigenvalue $\lambda_1=0$ as aforementioned.
 The product of the other $N-1$ eigenvalues $\lambda_\a$
of {\bf L} is known from graph theory \cite{Biggs,tw} to be
equal to $N$ times its spanning tree generating function with edge
weights $y_1, y_2, ..., y_N$.  Now the $N$
spanning trees are easily written down and as a result we obtain
\bea
\prod_{i=2}^N \lambda_\a
       &=& N(-j)^{N-1} \bigg( \frac 1 {y_1} +\frac 1 {y_2}+ \cdots + \frac 1 {y_N} \bigg)
                               y_1y_2 \cdots y_N \nonumber \\
                        &=&N(-j)^{N-1} (x_1 + x_2 + \cdots + x_N)/x_1x_2\cdots x_N.
\eea
 It follows that there exists another zero eigenvalue, and hence
a resonance, if $x_1 + x_2 + \cdots +x_N =0$.  This determines the resonance frequency $\omega$.

\bigskip
\noindent
{\it Example 3. A one-dimensional ring of N equal impedances}.

\medskip
In this example we consider $N$ equal impedances $z$ connected in a ring.  We have
\be
{\bf L} = y\,{\bf T}^{\rm per}_N, \quad {\bf L}^\dagger = y^*\, {\bf T}^{\rm per}_N,
\quad {\bf L}^\dagger\,{\bf L} =  |y|^2 \,\big({\bf T}^{\rm per}_N\big)^2
\ee
where $y=1/z$ and
\bea
{\bf T}_{N}^{\rm per}=\left(\begin{array}{ccccccc}
2&-1&0&\cdots&0&0&-1\\
-1&2&-1&\cdots&0&0&0\\
\vdots&\vdots&\vdots&\ddots&\vdots&\vdots&\vdots\\
0&0&0&\cdots&-1&2&-1\\
-1&0&0&\cdots&0&-1&2
\end{array}
\right) .\label{TNper}
\eea
  Thus
{\bf L} and ${\bf L}^\dagger\,{\bf L}$ all have the same eigenvectors.
  The eigenvalues and orthonormal eigenvectors
of ${\bf T}^{\rm per}_N$ are
\bea
\mu_n &=& 2[1-\cos (2n\pi/N) ]=4\cos^2 (n\pi/N)  \nonumber \\
\psi_n &=& \frac 1 {\sqrt N} \pmatrix{ 1\cr \omega^{n} \cr \omega^{2n} \cr \vdots \cr \omega^{(N-1)n}},
\quad n=0,1,...,N-1
\eea
where $\omega=e^{i 2\pi/N}$. The eigenvalues of ${\bf L}^\dagger\, {\bf L}$ are
\be
\s_n = |y|^2 \mu_n^2.
\ee
Since
\be
 \s_{N-n} =\s_n \,,
\ee
the corresponding eigenvectors are degenerate and we need to construct vectors
$u_{n1}$ and $u_{n2}$ for  $0<n<N/2$. For $N=$ even, however, the
eigenvalue $\s_{N/2}$ is non-degenerate and needs to be considered separately.

\medskip
For $0<n<N/2$ the degenerate eigenvectors
\be
\psi_n \quad {\rm and} \quad \psi_{N-n} = \psi_{n}^*\,
\ee
are not orthonormal.
Then we  construct linear combinations
  \bea
u_{n1} &=& \frac {\psi_n + \psi_{n}^*} {\sqrt 2}= \sqrt \frac 2 N
     \pmatrix{1 \cr \cos \frac {2n\pi} N \cr \cos \frac {4n\pi} N \cr\vdots
                         \cr \cos \frac {2(N-1)n\pi} N \cr} ,\nonumber \\
u_{n2} &=& \frac {\psi_n - \psi_{n}^*}{\sqrt 2 \, i} = \sqrt \frac 2 N
     \pmatrix{0 \cr \sin \frac {2n\pi} N \cr \sin \frac {4n\pi} N \cr\vdots
                         \cr \sin \frac {2(N-1)n\pi} N \cr}, \ \ n=1,2,\cdots,\bigg[\frac {N-1} 2\bigg],
\nonumber \\
  \eea
which are  orthonormal, where $[x]=$ the integral part of $x$.
The $u$'s are  eigenvectors of
${\bf L}^\dagger\, {\bf L}$ with the same eigenvalue $\s_n=|y|^2\mu_n^2$.
 For $N=$ even we have an additional non-degenerate eigenvector
\be
u_{N/2} = \frac 1 {\sqrt N} \pmatrix{\ \,1\cr -1\cr \ \,1\cr -1\cr \vdots \cr -1}.
\ee

\medskip
We next use (\ref{u}) to determine the phase factors $\theta_{n1}$ and $\theta_{n2}$.
Comparing the eigenvalue equation
\bea
&&{\bf L}\,u_{n1} = (y \mu_n)  u_{n1}\quad  {\rm with} \quad
{\bf L} \, u_{n1} = (|y| \mu_n) e^{i\theta_{n1}}\, u_{n1}^*, \nonumber \\
 \quad&&
{\bf L}\, u_{n2}   =(y \mu_n)  u_{n2}\quad  {\rm with} \quad
{\bf L}\, u_{n2} = (|y| \mu_n) e^{i\theta_{n2}}\, u_{n2}^*, \nonumber\\
{\rm and} && {\bf L} u_{N/2} =4(y) u_{N/2} \quad  {\rm with} \quad {\bf L} u_{N/2} = 4|y| e^{i\theta_{N/2}}
u_{N/2}^*
\eea
 we obtain
\be
\theta_{n1} =  \theta_{n2} = \theta_{N/2}= \theta,
\ee
where $\theta$ is given by $y=|y|\, e^{i\theta}$.

\medskip
We now use (\ref{Impedance}) to compute the impedance between nodes $p$ and $q$
to obtain
\bea
Z_{pq}&=& \frac 2 {Ny}  \sum_{n=1}^{[ \frac {N-1} {2} ] }
\frac 1 {  \mu_n}
\Bigg[ \bigg( \cos \frac {2np\pi} N -\cos \frac {2nq\pi} N\bigg)^2
- i^2   \bigg( \sin \frac {2np\pi} N -\sin \frac {2nq\pi} N\bigg)^2\Bigg]
\nonumber \\
&& + \, E
\eea
where $[x]$ denotes the integral part of $x$ and
\bea
E &=& \frac {1} {2Ny} \bigg[(-1)^p - (-1)^q \bigg]^2,\quad N={\rm even} \nonumber \\
  &=& 0 , \hskip 4.4cm N= {\rm odd}.
\eea
After some manipulation it is reduced to
\be
Z_{pq} = \frac z N \sum_{n=1} ^ {N-1} \frac
{ \Big|e^{i2 n p \pi/N} -e^{i2 n q \pi/N} \Big|^2} { 2[1- \cos (2n \pi/N)]} .
\ee
This expression has been evaluated in \cite{wu} with the result
\be
Z_{pq} = z \,|p-q| \bigg[1 - \frac {|p-q|} N \bigg],
\ee
which is the expected impedance of two impedances $|p-q|z$ and
$ (N - |p-q|)z$ connected in parallel as in a ring.
This completes the evaluation of $Z_{pq}$.

\bigskip
\noindent
{\it Example 4. Networks of inductances and capacitances.}

 \medskip
As an example of networks of inductances and capacitances, we consider an $M\times N$
array of nodes forming a rectangular net with free boundaries
as shown in Fig. 3.  The nodes are connected
by capacitances $C$ in the $M$ directions and inductances $L$ in the $N$ direction.

\medskip
The Laplacian of the network is \be {\bf L} = \big( j \omega C
\big)\,{\bf T}^{\rm free}_M \otimes {\rm I}_N
           - \Big(\frac j {\omega L}\Big)\, {\rm I}_M \otimes {\bf T}^{\rm free}_N
\ee
where ${\bf T}^{\rm free}_M$ is the $M\times M$ matrix
\be
{\bf T}_M^{\,\rm free}=\left(\begin{array}{ccccccc}
1&-1&0&\cdots&0&0&0\\
-1&2&-1&\cdots&0&0&0\\
 \vdots&\vdots&\vdots&\ddots&\vdots&\vdots&\vdots\\
0&0&0&\cdots&-1&2&-1\\
0&0&0&\cdots&0&-1&1
\end{array} \label{TNfree}
\right),
\ee
and ${\bf I}_N$ is the $N\times N$ identity matrix.
This gives
\be
 {\bf L}^*{\bf L} = \big( \omega C\big)^2 \,{\bf U}^{\rm free}_M \otimes {\bf I}_N
-2 \Big( \frac C L \Big)\, {\bf T}^{\rm free}_M \otimes {\bf T}^{\rm free}_N
     + \Big( \frac 1 {\omega L } \Big)^2\,  {\bf I}_M \otimes {\bf U}^{\rm free}_N
\ee
where ${\bf U}^{\rm free}_M $ is the $M\times M$ matrix
\be
{\bf U}^{\rm free}_M = \left(
\begin{array}{ccccccccc}
2 & -3 & 1 & 0 & 0 & \cdots & 0 & 0 & 0 \\
-3 & 6 & -4 & 1 & 0 & \cdots & 0 & 0 & 0 \\
1 & -4 & 6 & -4 & 1 & \cdots & 0 & 0 & 0 \\
0 & 1 & -4 & 6 & -4 & \cdots & 0 & 0 & 0 \\
0 & 0 & 1 & -4 & 6 & \cdots & 0 & 0 & 0 \\
\vdots & \vdots & \vdots & \vdots & \vdots & \ddots & \vdots & \vdots &
\vdots \\
0 & 0 & 0 & 0 & 0 & \cdots & -4 & 6 & -3 \\
0 & 0 & 0 & 0 & 0 & \cdots & 1 & -3 & 2
\end{array}
\right) .
\ee
Now ${\bf T}^{\rm free}_M $ has eigenvalues
\be
\lambda_m = 2(1-\cos \t_m)=4 \sin^2 (\t_m/2), \quad \t_m = \frac {m\pi} M
\ee
and eigenvector $\psi^{(M)} _m$ whose components are
\be
\psi _{mx}^{( M) }=\left\{
\begin{array}{cc}
\frac{1}{\sqrt{M}}, & m=0, {\rm \>\> for\>\> all \>\>}x, \\
\sqrt{\frac{2}{M}}\cos \Big( x+ \frac 1 2\Big) \t_m,  & m=1,2,\ldots
,M-1, {\rm \>\> for\>\> all\>\> }x.
\end{array}
\right.
\ee
It follows that ${\bf L}^*{\bf L}$ has eigenvectors
\be
\psi ^{\rm free}_{(m,n);(x,y)} = \psi _{mx}^{( M) } \psi _{ny}^{( N) }
\ee
and eigenvalues
\be
 \s_{mn} = 16 \Big( \omega C \, \sin^2\frac {\t_m} 2    -\frac 1 {\omega L}
\sin^2 \frac {\phi_n} 2 \Big)^2
\ee
where $\t_m = m\pi/M, \ \phi_n=n\pi/N$.
This gives
\bea
 \l_{mn} &=& 4j\Big[\omega C \sin^2 (\t_m/2)  -  \frac 1{\omega L} \sin^2 (\phi_n/2)
 \Big]\nonumber \\
         &=& \sqrt {\s_{mn}}\, e^{i\t_{mn}},\quad \t_{mn} = \pm\, \pi/ 2\, .
\eea
Since the vectors $\psi ^{\rm free}_{(m,n);(x,y)}$ are orthonormal and non-degenerate,
according to the Theorem
we can use these vectors in (\ref{Impedance}) to obtain
the impedance between nodes  $(x_1,   y_1)$ and
$(x_2,y_2)$.  This gives
\bea
&& Z^{\rm free}_{(x_1,y_1);(x_2,y_2)} =
 {\sum_{m=0}^{M-1}\sum_{n=0}^{N-1}}_{(m,n) \neq (0,0)}
 \frac { \Big( \psi^{\rm free} _{(m,n); (x_1,y_1)}- \psi^{\rm free} _{(m,n); (x_2,y_2)}\Big)^2} {\l_{mn}}
 \nonumber \\
&& \hskip 1cm = \frac {-j} {N\omega C} \big| x_1-x_2 \big| +\frac {j\omega L} M \big| y_1-y_2 \big|
+ \frac {2j} {MN} \nonumber \\
&&  \times \sum_{m=1}^{M-1}\sum_{n=1}^{N-1}
  \frac {\Big[ \cos \Big( x_1+ \frac 1 2\Big) \t_m \,\cos \Big( y_1+ \frac 1 2\Big) \phi_n
              -\cos \Big( x_2+ \frac 1 2\Big) \t_m \,\cos \Big( y_2+ \frac 1 2\Big) \phi_n \Big]^2}
        { - \omega C (1- \cos \t_m )  + \frac 1{\omega L} (1- \cos \phi_n ) }. \nonumber \\ \label{freebc}
  \eea

\medskip
As discussed in section 4, resonances occur at AC frequencies determined from $\l_{mn}=0$.
Thus, there are $(M-1)(N-1)$ distinct resonance frequencies given by
 \be
\omega_{mn} = \Bigg| \frac {\sin (n\pi / 2N)} {\sin (m\pi / 2M)}
\Bigg| \frac 1 {\sqrt {LC}}, \quad m=1,...,M-1;\, n=1,...,N-1.
\label{resonances} \ee A similar result can be found for an $M\times
N$ net with toroidal boundary conditions. However, due to the
degeneracy of eigenvalues, in that case there are
$[(M+1)/2][(N+1)/2]$ distinct resonance frequencies, where $[x]$ is
the integral part of $x$. It is of pertinent interest to note that a
network can become resonant at a spectrum of distinct frequencies,
and these resonances occur in the effective impedances between {\it any}
two nodes.

\medskip
In the limit of $M,N \to \infty$,  (\ref{resonances}) becomes continuous indicating that the
network is resonant at all frequencies. This is verified by replacing the summations by
integrals in
(\ref{freebc}) to yield the effective impedance between
two nodes $(x_1,y_1)$ and $(x_2,y_2)$
\bea
&&Z^{\infty}_{(x_1,y_1);(x_2,y_2)} \nonumber \\
 && =
\frac {j}  {4\pi^2}\int_0^{2\pi}d\theta\int_0^{2\pi}d\phi
 \Bigg[ \frac{1-\cos [(x_1-x_2) \theta]\, \cos [(y_1-y_2) \phi]}
          {-\omega C (1- \cos \t)  +  \frac 1{\omega L} (1- \cos \phi )} \Bigg] ,
\label{infinite}
\eea
which diverges
logarithmically.\footnote{Detailed steps leading to  (\ref{freebc}) and (\ref{infinite})
can be found in Eqs. (37) and (40) of \cite{wu}.}

\section{Summary}
We have presented a formulation of impedance networks which permits the evaluation of the effective impedance
between arbitrary two nodes.  The resulting expression is
 (\ref{Impedance}) where $u_\a$ and $\l_a$ are those given in ({\ref{modZeigen}).
In the case of reactance networks, our analysis indicates that resonances occur
at AC frequencies $\omega$ determined by the vanishing of $\l_a$.
This curious result suggests the possibility of practical applications
of our formulation to resonant circuits.

\section*{Acknowledgment}

\medskip
This work was initiated while both authors were at the National Center of
Theoretical Sciences (NCTS) in Taipei.  The support of the NCTS is
gratefully acknowledged.
 Work of WJT has been supported in part by National Science Council
grant NSC 94-2112-M-032-008. We are grateful to J. M. Luck for calling
our attention to \cite{clerc,clerc1} and M. L. Glasser for pointing us
to \cite{asad}.

\newpage
\begin{figure}[h]
\includegraphics{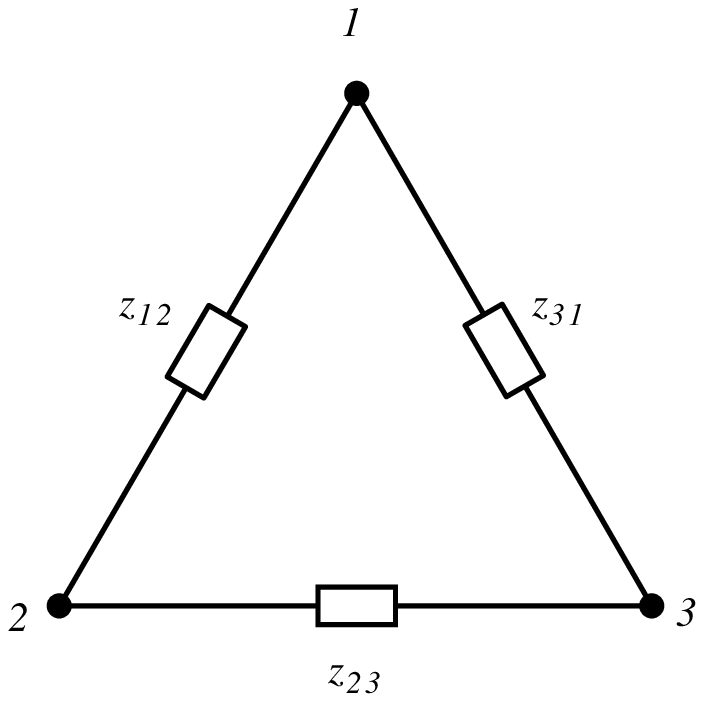} \vskip 11.5cm
\end{figure}
\noindent Fig.\ 1~~  An example of three impedances in a ring.

\newpage
\begin{figure}[h]
\includegraphics{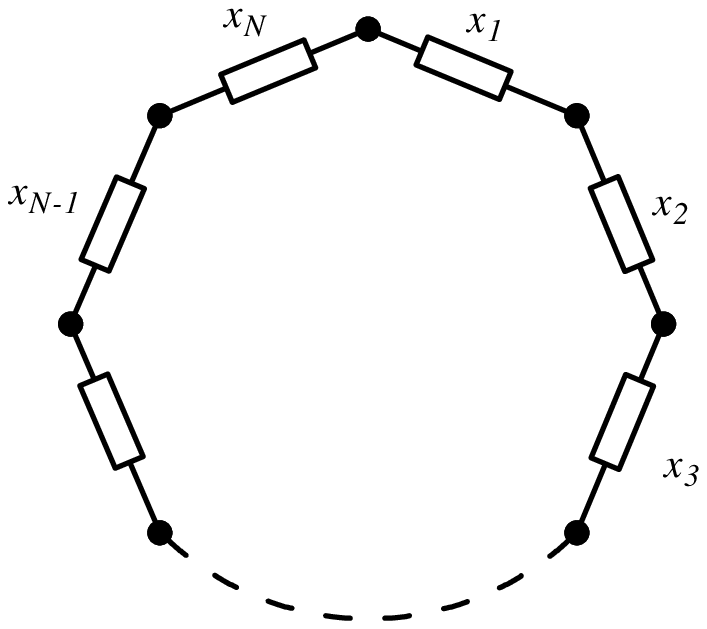} \vskip 11.5cm
\end{figure}
\noindent Fig.\ 2~~  A ring of $N$  reactances.

\newpage
\begin{figure}[h]
\includegraphics{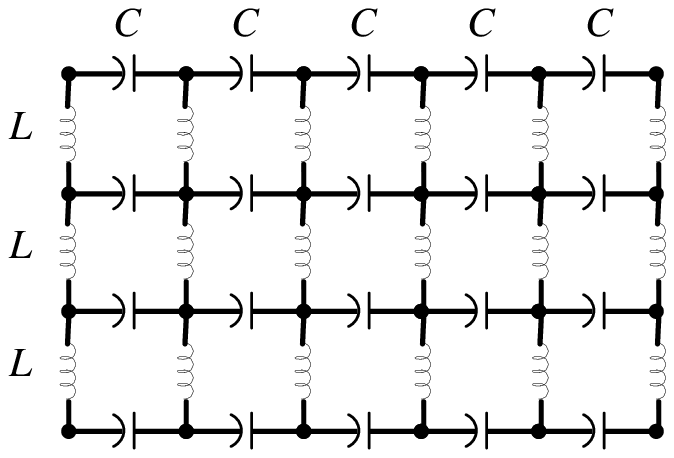} \vskip 11.5cm
\end{figure}
\noindent Fig.\ 3~~  A $6 \times 4$ network of capacitances $C$ and
inductances $L$.

 \end{document}